\documentclass[twocolumn]{article}
\usepackage{amssymb}

\def\be{\begin{equation}}
\def\ee{\end{equation}}
\def\bea{\begin{eqnarray}}
\def\eea{\end{eqnarray}}
\def\Q{{\cal Q}}
\def\P{{\cal P}}
\def\A{{\cal A}}

\def\th{^{\mbox{\scriptsize th}}}

\def\part{{\cal P}_{[s]}}
\def\Om{\Omega_{[s]}}
\def\Omz{\Omega_{[0]}}
\def\cpart{\overline{{\cal P}}_{[s]}}
\newcommand{\w}[1]{(\ref{#1})}

\begin{document}

\title{Multipartite Secret Correlations and Bound Information}
\author{Llu\'\i s Masanes\footnote{School of Mathematics, University of Bristol,
Bristol BS8 1TW, United Kingdom} \footnote{Dept. d'Estructura i
Constituents de la Mat\`eria, Univ. de Barcelona, 08028 Barcelona,
Spain} \hspace{.02cm} and Antonio Ac\'\i n\footnote{ICFO-Institut
de Ci\`encies Fot\`oniques, Jordi Girona 29, Edifici Nexus II,
08034 Barcelona, Spain}}

\date{\today }
\maketitle
\begin{abstract}
We consider the problem of secret key extraction when $n$ honest
parties and an eavesdropper share correlated information. We
present a family of probability distributions and give the full
characterization of its distillation properties. This formalism
allows us to design a rich variety of cryptographic scenarios. In
particular, we provide examples of multipartite probability
distributions containing non-distillable secret correlations, also
known as bound information.
\end{abstract}

\section{Introduction}

Many cryptographic applications nowadays are based on {\em
computational security}. In this type of protocols, the security
is based on two assumptions: (i) the computational capabilities of
an eavesdropper are bounded and (ii) a conjecture on the
computational complexity of some mathematical problems. The advent
of quantum computing however sheds doubts on the medium-term
applicability of these schemes. Indeed, Shor's algorithm
\cite{Shor} will allow an eavesdropper provided with a quantum
computer breaking many of the now commonly used schemes, such as
RSA.

There is a second type of security which is clearly the strongest
one: {\em information-theoretic security}. This type of protocols
are secure against attacks using unlimited resources, since the
security is simply guaranteed by known results of Information
Theory. The first step in this direction was already given in 1949
by Shannon \cite{Shannon}: such a level of security could only be
attained by honest parties initially sharing a secure secret key.
An example of a completely secure way of information encryption is
given by the one-time pad \cite{Vernam}: in this scheme the honest
parties share a private key. The message is summed (XOR) bitwise
with the common key and sent through the insecure public channel.
The receivers owning the key can read the sent information
performing a second bitwise XOR, while no information on the
message is accessible to anybody with no access to the key. It
turns out however that (i) this protocol works only when the
parties willing to interchange the information share a private key
of the same length as the message to be encrypted and (ii) the key
cannot be reused. Moreover, the following question arises in a
natural way: how is the key generated? It was later shown by
Maurer that a secure key cannot be generated from nothing
\cite{Maurer}. More precisely, the honest parties cannot establish
a key by a protocol consisting of local operations and public
communication (LOPC). Therefore, a necessary requirement for
key-agreement is that the honest parties share prior correlations
that are partially secret.

These pessimistic statements were somehow relativized in
\cite{Maurer,MW}, where it was proven that an arbitrary weak level
of correlation and privacy can be in some cases sufficient for
generating a key. Furthermore, Quantum Cryptography protocols
\cite{BB84,review} have been shown to provide an efficient way for
establishing these initial partially secret correlations. Indeed,
it is a crucial problem in most of Quantum Cryptography protocols
how to transform into a perfect secret key the noisy and partially
secret data distributed among the honest parties through quantum
channels. As said, this key will later be consumed for sending
private information by means of one-time pad.

In this work, we study the inter-conversion among different kinds
of secret correlations in a multipartite scenario, where $n$
honest parties and an eavesdropper have access to common
information. More precisely, each party, including the
eavesdropper, has many realizations of a random variable. These
$n+1$ random variables are correlated through a known probability
distribution. Since the secrecy content of these correlated data
is non-increasing under LOPC, this set of transformations is
considered as a free resource. That is, the honest parties are
allowed to perform any local operation on their data and to
communicate through a public, but authenticated, channel. Given an
initial probability distribution $P$, we focus on two questions:
(i) can $P$ be generated by LOPC? and (ii) can perfect secret bits
be extracted from $P$ by LOPC?

A family of probability distributions is introduced, allowing the
construction of several examples (see below) with a huge variety
of distillation properties. In particular, for each probability
distribution we can answer the previous two questions (i) and
(ii). Note that we do not consider the problem of how these
correlations are generated, that is, they will appear as an
initially given resource. However, there have been proposed
different ways of establishing partially secret correlations, such
as the satellite model by Maurer \cite{Maurer}, or Quantum
Cryptography \cite{BB84}. The analyzed techniques can be used to
prove the existence and activation of {\em bound information}, a
cryptographic analog of bound entanglement, first conjectured in
\cite{GW} (see also \cite{GRW}). We finally discuss the connection
of these techniques with previous results on entanglement
transformations in Quantum Information Theory.

\subsection{Examples}
\label{exsec}

In this section we present some examples of multipartite secret
correlations showing interesting distillability properties. These
probability distributions are explicitly constructed in section
\ref{exconstr}. In the following examples, we consider $n$
separated parties $\{\A_1,\A_2,\ldots, \A_n\}$. Throughout the
article, whenever we say that a subset of $k$ parties are
together, or form a group, we mean that they can perform
$k$-partite joint secret operations. This can be done by meeting
at the same place, or by sharing a sufficiently large $k$-partite
secret key.

{\bf Example 1.} $n$ honest parties can distill a secret key if at
least 70\% of them cooperate in the protocol, independent of the
fact that they join or not (the choice of 70\% is arbitrary).

{\bf Example 2.} Probability distribution of $n$ honest parties
where distillation is possible if, and only if, the cooperating
parties join in groups of at least $k$ people, independently of
how many parties participate in the distillation protocol.

The previous two examples can be considered as elementary
conditions on distillation. In the following two examples we
combine them with different logical clauses (AND and OR), in order
to obtain more sophisticated distillation scenarios.

{\bf Example 3.} $n$ honest parties can obtain a secret key if,
and only if, at least 70\% of them cooperate in the protocol, AND,
they join in groups of at least $k$ people.

{\bf Example 4.} $n$ honest parties can generate a secure key if,
and only if, at least 70\% of them cooperate in the protocol, OR,
the cooperating parties join in groups of at least $k$ people, or
both.

In the previous examples all the parties played the same role. In
the following some specific parties have a different status, that
is, the possibility of distillation may depend on their actions.

{\bf Example 5.} Distribution of $n$ honest parties where
distillation is possible if, and only if, the parties $\A_i \A_j$
participate in the protocol and remain together, independently of
how many others cooperate and how they distribute in groups. The
same can be done but imposing that parties $\A_i \A_j$ must remain
separated.

It is now clear that one can design probability distributions
showing unlimited intricate distillation properties.

\section{Bipartite secret correlations}

In 1993, Maurer introduced the information-theoretic key-agreement
model, generalizing previous ideas by Wyner \cite{Wyner} and
Csisz\'ar and K\"orner \cite{CK}. In his original formulation, two
honest parties (Alice and Bob) are connected by an authenticated
but otherwise insecure classical communication channel.
Additionally, each party
---including Eve--- has access to correlated information given by
repeated realizations of the random variables $A$, $B$ and $E$
(possessed by Alice, Bob and Eve respectively), jointly
distributed according to $P(A,B,E)$. From now on, we denote by the
same symbol, $X$, a random variable, $X$, as well as the value it
can take, $x$, e.g. $P(X)=P_X(x)$. The goal for Alice and Bob is
to obtain a common string of random bits for which Eve has
virtually no information, i.e. a secret key. The maximal amount of
secret key bits that can be asymptotically extracted per
realization of $(A,B)$ used, is called the secret-key rate,
denoted by $S(A:B\parallel E)$ or $S$. More precisely, this
quantity is defined as the largest real number such that for all
$\epsilon>0$, one can find an integer $N_0$ and a two-way
communication protocol for Alice and Bob transforming $N\geq N_0$
realizations of $A$ and $B$ into random variables $S_A$ and $S_B$
satisfying
  \bea
  P[S_A=S_B=X]&>&1-\epsilon \nonumber\\
  H(X)=\log(|X|)&\geq& (R-\epsilon)N \nonumber\\
  I(X:CE^N)&<&\epsilon ,
  \eea
where $X$ is another random variable and $C$ denotes the
communication exchanged during the protocol. Therefore, the
secret-key rate quantifies the amount of secret-key bits
extractable from a probability distribution.

More recently, another measure for the secrecy content of
$P(A,B,E)$, the so-called information of formation
$I_{\mbox{\scriptsize form}}(A:B|E)$, has been introduced in
\cite{RW}. Intuitively, it can be understood as the minimal number
of secret-key bits asymptotically needed to generate each
independent realization of $(A,B)$ ---distributed according to
$P(A,B)$---, such that the information about $(A,B)$ contained in
the messages exchanged through the public channel, $C$, is at most
equal to the information in $E$. 
More precisely, $I_{\mbox{\scriptsize form}}$ is defined as the
infimum over all numbers $R\geq 0$ such that for all $\epsilon>0$
there exists an integer $N$, and a protocol with communication $C$
that, with probability $1-\epsilon$, allows Alice and Bob, knowing
the same random $\lceil RN\rceil$-bit string $X$, to compute $A^N$
and $B^N$ such that
  \be
  \label{infform}
  P(A^N,B^N,C)=\sum_{E^N} \left[ P(A,B,E) \right]^N P(C|E^N) ,
  \ee
where $P(C|E^N)$ defines a channel \cite{RW}. According to this
definition, we say that a probability distribution $P$ can be
established by LOPC if, and only if, $I_{\mbox{\scriptsize
form}}=0$. Note that this statement does not mean that the result
of the corresponding LOPC formation protocol is necessarily $P$,
but it is a distribution $P'$ at least as good as $P$ from Alice
and Bob's point of view. More concretely, $P'$ can be obtained
from $P$ by processing Eve's information, in the sense of Eq.
(\ref{infform}).

Information of formation and secret-key rate are two measures of
the secrecy content of a probability distribution with a clear
operational meaning: $I_{\mbox{\scriptsize form}}$ quantifies the
amount of secret-key bits required for the formation of
$P(A,B,E)$, while $S$ specifies the amount of secret bits
extractable from $P(A,B,E)$.

A useful upper bound for $S$ is given by the so-called intrinsic
information, introduced in \cite{MW}. This quantity, denoted by
$I(A:B\downarrow E)$ or more briefly $I_\downarrow$, will play a
significant role in the proof of our results. The intrinsic
information between $A$ and $B$ given $E$ is defined as: \be
  I(A:B\downarrow E)\ =\ \min_{E\rightarrow \tilde{E}}\ I(A:B|\tilde{E})\ ,
  \label{intrinf}
\ee where the minimization runs over all possible stochastic maps
$P_{\tilde E|E}$ defining a new random variable $\tilde{E}$. The
quantity $I(A:B|E)$ is the mutual information between $A$ and $B$
conditioned on $E$. It can be written as \be
  I(A:B|E)=H(A,E)+H(B,E)-H(A,B,E)-H(E) \ ,
\ee where $H(X)$ is the Shannon entropy of the random variable
$X$. The intrinsic information also gives a lower bound for the
information of formation \cite{RW}, thus
\begin{equation}
\label{upb} S\leq I_\downarrow\leq I_{\mbox{\scriptsize form}}\ .
\end{equation}

\section{Multipartite secret correlations}

The generalization of Maurer's formulation to the multipartite
scenario is straightforward. Consider a set of $n$ honest parties
$\Q =\{ \A_1, \A_2, \ldots \A_n \}$ connected by a broadcast
public communication channel which is totally accessible to the
eavesdropper but which is tamper-proof. Each of the parties
(including Eve) has access to the correlated information contained
in many realizations of its corresponding random variable. We
denote by $A_i$ the random variable corresponding to party $\A_i$.
Eve's random variable is also denoted by $E$. In the whole paper
curly capital letters refer to parties and sets of parties, while
normal capital letters refer to random variables. In this
scenario, general secret correlations are represented by
probability distributions of the form $P(A_1,\dots ,A_n,E)$. That
is, all random variables in each realization are correlated
according to $P$, and each realization is independent of the
others.

One possible goal for the honest parties is to obtain an
$n$-partite secret key. Sometimes this is not possible, but still,
a subset of $m$ parties (with $1<m<n$) can get an $m$-partite key.
Therefore, there are many different senses in which a distribution
$P$ is (or is not) distillable. In order to get rid of such
ambiguity we choose the strongest definition of
non-distillability. We say that a distribution $P$ is
non-distillable if there does not exist any pair of parties,
capable of obtaining a secret key by LOPC, even with the help of
the others. For similar reasons, in the multipartite scenario
there may be many ways of defining the secret-key rate. But, in
this paper we only use the secret-key rate in bipartite
situations, where the definition is unique. In general,
considering bipartite splittings of the parties will prove to be a
very useful tool for obtaining necessary conditions in the
multipartite scenario.

\subsection{Bipartite splittings}

We denote by $\P$ any subset of $\Q$, and by $\overline{\P}$ its
complement (the set of all elements in $\Q$ not belonging to
$\P$). Each bipartition of $\Q$ can be specified by giving one of
the halves, say $\P$.

The following two lemmas concerning any $n$-party distribution
refer to their distillation and formation properties.

\bigskip

{\bf Lemma 1:} {\em A necessary condition for obtaining an
$n$-partite secret key is that: for all bipartitions $\P$ of $\Q$,
when all parties within each half are together, a bipartite secret
key between $\P$ and $\overline{\P}$ can be obtained.}

\bigskip

{\em Proof:} Suppose the distribution can be distilled into an
$n$-partite secret key. The same must hold when some of the
parties are together. In particular, a bipartite key between the
groups $\P$ and $\overline{\P}$ can be obtained, for any $\P$.
Therefore, the last is a necessary condition. $\Box$

\bigskip

{\bf Lemma 2:} {\em A necessary condition for the correlations
specified by $P$ being generated by LOPC is that: for all
bipartitions $\P$ of $\Q$, when all parties within each half are
together, the resulting bipartite distribution can be generated by
LOPC.}

\bigskip

{\em Proof:} Suppose the distribution can be generated using LOPC
by the $n$ honest parties. The same must hold when some of the
parties are together, in particular, for the bipartite splitting
$\P$ and $\overline{\P}$. Therefore, the last is a necessary
condition. $\Box$

\bigskip

\section{A family of multipartite probability distributions}

In this section we present a family of probability distributions,
denoted by $P_\Omega$, exhibiting a variety of distillation
properties. The examples described in section \ref{exsec} are
particular instances of this family.

\subsection{Notation and definitions}

From now on, we restrict the random variables of the honest
parties $A_1,\ldots ,A_n$ to take the values $0,1$. Eve's random
variable $E$ can however have a wider range. In the remainder,
unless explicitly mentioned, quantities between square brackets
$[s]$ are to be understood as $(n-1)$-bit strings. That is, we
associate with each integer $s\in \{0,1, \ldots2^{n-1}-1\}$ the
$(n-1)$-bit string corresponding to its binary expansion, and
denote this by $[s]$. We denote by $[\bar{s}]$ the string where
all bits have the opposite value than in $[s]$. As an instance,
suppose $n-1=3$, we have that $[2]=010$ and $[\bar{2}]=101$. We
will use bit strings $[s]$ to label the outcome of the first
$(n-1)$ variables $A_1,\ldots,A_{n-1}$; for example $A_1\ldots
A_{n-1}=[s]$.

In what follows, we also use bit strings $[s]$ to specify
bipartitions of the set of $n$ parties $\Q$. The subset $\part
\subset \Q$ is defined in this way: $A_i \in
\part$ if the $i\th$ bit of $[s]$ is one. Notice that $A_n$
always belongs to $\cpart$. In this way, we associate with each
$(n-1)$-bit string $[s]$ a bipartition of $\Q$. As an example
suppose $\Q=\{ A_1,A_2,A_3 \}$, the string $01$ corresponds to the
bipartition $(A_2)-(A_1A_3)$, and, $00$ corresponds to the trivial
bipartition $()-(A_1 A_2 A_3)$.

Let us denote by $P_\Omega (A_1,\ldots A_n,E)$ the following
family of probability distributions.
\begin{center}
\begin{tabular}{|cc|c|c|} \hline
  $A_1\ldots A_{n-1}$ &$A_n$ &$E$ &$P_\Omega$ \\
  \hline\hline
  $[0]$ &0 &$[0]0$ or $[\bar{0}]1$ &$\Omega_{[0]}$ \\
  $[\bar{0}]$ &1 &$[0]0$ or $[\bar{0}]1$ &$\Omega_{[0]}$ \\ \hline
  $[1]$ &0 &$[1]\,0$ &$\Omega_{[1]}$ \\
  $[\bar{1}]$ &1 &$[\bar{1}]\,1$ &$\Omega_{[1]}$ \\
  &&&\\
  \vdots &\vdots &\vdots &\vdots \\
  &&&\\
  $[s]$ &0 &$[s]\,0$ &$\Om$ \\
  $[\bar{s}]$ &1 &$[\bar{s}]\,1$ &$\Om$ \\
  &&&\\
  \vdots &\vdots &\vdots &\vdots \\
  &&&\\
  $[2^{n-1}-1]$ &0 &$[2^{n-1}-1]\,0$ &$\Omega_{[2^{n-1}-1]}$ \\
  $[ \overline{2^{n-1}-1} ]$ &1 &$[\overline{2^{n-1}-1}]\,1$ &$\Omega_{[2^{n-1}-1]}$ \\ \hline
\end{tabular}
\vspace{10pt}
\end{center}
In this table, each row corresponds to a different event. For
example, in the first row event, the obtained outcomes are
$(A_1\ldots A_n)=(0\ldots 0)$ and $E=``[0]0 \mbox{ or }
[\bar{0}]1"$, and this happens with probability $\Omega_{[0]}$. As
can be seen, the $2^n$ events are grouped in equiprobable pairs.
Notice that Eve always knows the value of $A_1 \ldots A_n$, except
in the first two events, where she obtains the outcome ``$[0]0
\mbox{ or } [\bar{0}]1$" independently of which is the actual one.
The parameters of $P_\Omega$ are the positive numbers
$\Omega_{[0]},\ldots,\Omega_{[2^{n-1}-1]}$, only constrained by
the normalization condition:
  \be
  \sum_{s=0}^{2^{n-1}-1} \Om =\frac{1}{2}\ .
  \label{norm}
  \ee
A simple example of $P_{\Omega}$ can be found in section
\ref{binfpr}.

\subsection{Bipartitions}

Let us study the bipartite properties of $P_\Omega$. In the
following lemmas, it is assumed that all parties within each half,
$\part$ and $\cpart$, are together.

\bigskip

{\bf Lemma 3:} {\em A bipartite secret key between the parts
$\part$ and $\cpart$ can be obtained if, and only if, $\Om <
\Omega_{[0]}$.}

\bigskip

{\em Proof:} To prove the {\em if} statement we only have to
provide an explicit distillation protocol. This protocol has two
steps. In the first step, the honest parties discard all
realizations of $P_\Omega$ in which not all the variables within
each half ($\part$ and $\cpart$) have the same value. This
operation only filters the following events:
  \be
  (A_1 \ldots A_{n-1}\, A_n) =
  [0]0,\ [\bar{0}]1,\ [s]0,\ [\bar{s}]1.
  \ee
The filtered probability distribution is, up to normalization:
\begin{center}
\begin{tabular}{|cc|c|c|} \hline
  $P_{[s]}$ &$\overline{P}_{[s]}$ &$E$ &$P_{\Omega|\mbox{\scriptsize filtered}}$ \\
  \hline\hline
  0 &0 &$[0]0$ or $[\bar{0}]1$ &$\Omz$ \\
  1 &1 &$[0]0$ or $[\bar{0}]1$ &$\Omz$ \\ \hline
  1 &0 &$[s]0$ &$\Om$ \\
  0 &1 &$[\bar{s}]1$ &$\Om$ \\
\hline
\end{tabular}
\end{center}
Where $P_{[s]}=A_i$ for all $i$ such that $\A_i \in
\part$, and analogously for $\overline{P}_{[s]}$ and $\cpart$.
Notice that this is well-defined because all parties in
$\part$($\cpart$) have obtained the same outcome. The second step
is the repeated code protocol, explained in the Appendix. There,
it is shown that this protocol generates a secret key if $\Omz
> \Om$, as we wanted to prove.

The {\em only if} part can be proven by showing that the intrinsic
information of this partition is zero, $\, I(\P_{[s]}
:\overline{\P}_{[s]} \downarrow E)=0$, when $\Omz  < \Om$. For
doing so, we perform the following stochastic map
$E\rightarrow\tilde{E}$: If $E$ is equal to $[s]\, 0$ or
$[\bar{s}]\, 1$, we assign $\tilde{E}:=$ ``$[0]0$ or $[\bar{0}]1$"
with probability $\Omz/\Om$, and, with probability $1-\Omz/\Om$ we
assign $\tilde{E}:=E$. In the rest of the cases we also assign
$\tilde{E}:=E$. It is easy to check that $I(\P_{[s]}\!
:\overline{\P}_{[s]} |\tilde{E})=0$, which ensures that
$I(\P_{[s]} :\overline{\P}_{[s]} \downarrow E)=0$. Now, the upper
bound \w{upb} implies that the secret-key rate must be also zero.
In other words, we have that when $\Omz < \Om$
  \be
  S(\P_{[s]} :\overline{\P}_{[s]}\|E)=0,
  \ee
which completes the proof. $\Box$

Lemma 3 provides a tool for designing distributions with involved
distillation properties. Suppose that, in order to distill a
secret key, the $n$ parties join in two groups according to the
bipartition $\part$. Now, we can choose in which of these
bipartitions distillation will be possible, and in which not. We
set $\Om=0$, if we allow the parties to obtain a secret key when
arranged according to $\part$. Notice that for non-trivial
bipartitions $[s]\neq [0]$. We set $\Om=\Omz$, if we forbid
distillation when the parties are arranged according to $\part$.
Finally, we set $\Omz$ such that the normalization condition
\w{norm} is satisfied. Notice that we have as many free parameters
as there are possible bipartitions.

\subsection{Multipartitions}

Lemma 3 tells us how to construct probability distributions
$P_\Omega$, choosing independently which bipartite splits permit
secret key extraction, and which do not. Next, we generalize Lemma
3 by considering situations in which the $n$ parties are joined in
more than two groups. Of course, this includes the case where the
$n$ parties are all separated. Let us introduce some notation
first.

An $m$-partition of $\Q$ is given by $m$ disjoint subsets
$\Q_1,\ldots ,\Q_m \subset \Q$ such that $\Q_1 \cup \cdots \cup
\Q_m = \Q$. As before, we consider that the parties within each
subset $\Q_i$ are together. We use $Q_i$ to denote the binary
variable associated with ``party" $\Q_i$.

\bigskip

{\bf Lemma 4:} {\em Consider an $m$-partition of $\Q$,
$\{\Q_1,\ldots ,\Q_m\}$. An $m$-partite secret key among these
groups of parties can be obtained if, and only if, for each bit
string $[s]$ such that its corresponding bipartition $\P_{[s]}$
does not split any set $\Q_1,\ldots ,\Q_m$, the inequality $\Om <
\Omz$ holds.}

\bigskip

{\em Proof:} The {\em only if} assertion is just Lemma 1, but
using the equivalence of Lemma 3. Let us prove the {\em if} part
by giving a protocol which is a generalization of the one given in
the proof of Lemma 3. First, the honest parties discard all
realizations of $P_\Omega$ in which there is at least one subset
$\Q_i$ containing variables with different values. Or
equivalently, they reject all events $A_1\ldots A_{n}=[s]0,\
[\bar{s}]1$ such that its associated bipartition $\part$ splits at
least one subset $\Q_i$. As usual, $Q_i=A_i$ for all $i$ such that
$\A_i\in\Q_i$. After this filtering operation the probability
distribution is, up to normalization:
\begin{center}
\begin{tabular}{|cc|c|c|} \hline
  $Q_1\ldots Q_m$ &$E$ &$P_{\Omega|\mbox{\scriptsize filtered}}$ \\
  \hline\hline
  $[0]0$ &$[0]0$ or $[\bar{0}]1$ &$\Omz$ \\
  $[\bar{0}]1$ &$[0]0$ or $[\bar{0}]1$ &$\Omz$ \\ \hline
  \vdots &\vdots &\vdots\\
  &&\\
  $[s]0$ &$[s]\,0$ &$\Omega_{[s]}$ \\
  $[\bar{s}]1$ &$[\bar{s}]\,1$ &$\Omega_{[\bar{s}]}$ \\
  &&\\
  \vdots &\vdots &\vdots \\ \hline
\end{tabular}
\vspace{10pt}
\end{center}
Notice that in the first column, we specify the value of the
$m$-bit string $Q_1\ldots Q_m$ with an $n$-bit string, say $[s]0$.
This is well defined if we recall that, in all filtered events,
the bits in $[s]0$ associated with the parties belonging to
$\Q_i$, have the same value, and this value is the one assigned to
the variable $\Q_i$. Now, the $m$ parties perform the repeated
code protocol to $P_{\Omega|\mbox{\scriptsize filtered}}$. In the
Appendix it is shown that this protocol works if the condition of
Lemma 4 holds. $\Box$

\subsection{Non-cooperating parties}

It is clear that a single party, say $\A_i$, can always prevent
the others from obtaining a secret key. To do so, she only has to
make public the value of $A_i$ in each realization of $P_\Omega$.
After this procedure, Eve will know the value of each variable in
the two events where all variables are equal: $(A_1\ldots
A_n)=(0\ldots 0),(1\ldots 1)$. In the rest of events, Eve already
knew the value of each variable. Therefore, by a non-cooperating
party we do not mean a party who is against the others, but one
that does not want to be involved in the distillation protocol. In
this section, we generalize Lemma 4 by considering the presence of
non-cooperating parties. Let us first, introduce some notation.

In what follows, when referring to the $m$ disjoint subsets
$\Q_1,\ldots, \Q_m \subset \Q$, we do not demand that they satisfy
$\Q_1 \cup \cdots \cup \Q_m = \Q$. In other words, they don't have
to be an $m$-partition of $\Q$. It is understood, that the parties
not belonging to $\Q_1 \cup \cdots \cup \Q_m$ do not participate
in the protocol. Within all this section, primed quantities
between square brackets $[z']$ have to be understood as
$(m-1)$-bit strings. That is, we associate with each integer
$z\in\{0,1,\ldots,2^{m-1}-1\}$ the $(m-1)$-bit string
corresponding to its binary expansion, denoted by $[z']$. As in
the rest of the paper, unprimed integers between square brackets
mean $(n-1)$-bit strings. We also denote by $[\bar{z}']$ the
$(m-1)$-bit string where each bit has the opposite value than in
$[z']$. Following the analogy, $\P_{[z']}$ is a subset of
$\{\Q_1,\ldots ,\Q_m\}$ defined with the same convention as
$\part$. That is, $\Q_i$ belongs to $\P_{[z']}$ if the $i\th$ bit
of $[z']$ has the value one. We define $\bar{\P}_{[z']}$
analogously, which always contains $\Q_m$. We also use $\P_{[z']}$
to denote bipartitions of $\{Q_1,\ldots,\Q_m\}$. Additionally, we
associate with each bipartition of $\{Q_1,\ldots,\Q_m\}$ some
bipartitions of $\Q$, in the following way. We say that $\part$ is
associated with $\P_{[z']}$ if $\part$ contains all parties
belonging to the subsets $\Q_i$ such that $\Q_i\in \P_{[z']}$,
and, does not contain any party belonging to the subsets $\Q_i$
such that $\Q_i\in \bar{\P}_{[z']}$. Notice that the
non-cooperating parties, the ones not belonging to
$\Q_1\cup\cdots\cup\Q_m$, may or may not belong to $\part$.
Therefore, there can be many $\part$ associated with one
$\P_{[z']}$. We also extend this relation to bit strings in a
natural way: we say $[s]\sim [z']$ if $\part$ is associated with
$\P_{[z']}$.

\bigskip

{\bf Theorem 5:} {\em An $m$-partite secret key among the groups
of parties $\Q_1,\ldots,\Q_m$ can be obtained if, and only if, for
each bipartition of these $m$ groups $\P_{[z']}$, the inequality
  \be
  \sum_{\forall [s]\sim [z']} \Omega_{[s]} < \Omz
  \label{non-coop}
  \ee
holds.}

\bigskip

{\em Proof:} As in the previous cases, we prove the {\em if}
assertion by giving a protocol that works under the stated
conditions. The usual protocol is readily generalized to fit this
case: The cooperating honest parties discard all realizations of
$P_\Omega$ for which there is at least one group $\Q_i$, in which
not all the variables are equal. Or equivalently, they reject all
events $A_1\ldots A_n=[s]0,\ [\bar{s}]1$ such that, its
corresponding bipartition $\part$ splits at least one subset
$\Q_i$.  Notice that in the filtered events, the non-cooperating
parties' variables can have any value. After this filtering, the
probability distribution is, up to normalization:
\begin{center}
\begin{tabular}{|cc|c|c|} \hline
  $Q_1\ldots Q_{m-1}$ &$Q_m$ &$E$ &$P_{\Omega|\mbox{\scriptsize filtered}}$ \\
  \hline\hline
  $[0']$       &$0$ &$[0']0$ or $[\bar{0}']1$ &$\Omz$ \\
  $[\bar{0}']$ &$1$ &$[0']0$ or $[\bar{0}']1$ &$\Omz$ \\ \hline
  $[1']$ &$0$       &$[1']\,0$       &$\sum_{\forall [s]\sim [1']} \Omega_{[s]}$ \\
  $[\bar{1}']$ &$1$ &$[\bar{1}']\,1$ &$\sum_{\forall [s]\sim [1']} \Omega_{[s]}$ \\
  &&&\\
  \vdots &\vdots &\vdots &\vdots \\
  &&&\\
  $[z']$ &$0$       &$[z']\,0$       &$\sum_{\forall [s]\sim [z']} \Omega_{[s]}$ \\
  $[\bar{z}']$ &$1$ &$[\bar{z}']\,1$ &$\sum_{\forall [s]\sim [z']} \Omega_{[s]}$ \\
  &&&\\
  \vdots &\vdots &\vdots &\vdots \\
  &&&\\
 \hline
\end{tabular}
\vspace{10pt}
\end{center}
As usual, $Q_i=A_i$ for all $i$ such that $\A_i\in\Q_i$. As shown
in the Appendix, the repeated code protocol works with
$P_{\Omega|\mbox{\scriptsize filtered}}$ if, for all $[z']$,
condition \w{non-coop} holds. To prove the {\em only if} part, let
us suppose that there exists at least one string $[z_0']$ such
that \w{non-coop} is not satisfied. According to Lemma 1, when the
groups $\Q_1,\ldots ,\Q_m$ can distill an $m$-partite secret key,
a bipartite key is also obtainable when the $m$ groups are joined
in just two groups. This must hold for any bipartition of
$\{\Q_1,\ldots \Q_m\}$, say $[z_0']$. Let us see that this is
impossible when
  \be
  \sum_{\forall [s]\sim [z_0']} \Omega_{[s]} \geq \Omz
  \ee
holds. As in the proof of Lemma 3, we show that the secret-key
rate between $\P_{[z_0']}$ and $\bar{\P}_{[z_0']}$ is zero, by
computing the intrinsic information between these two parts. To do
so, we perform a similar stochastic map $E\rightarrow\tilde{E}$:
If $E=[z']0$ or $E=[\bar{z}']1$ we assign $\tilde{E}=$``$[0]0$ or
$[\bar{0}]1$" with probability $\Omz/\sum_{\forall [s]\sim [z']}
\Omega_{[s]}$. In the rest of the cases $\tilde{E}=E$. It is easy
to check that
  \be
  I(P_{[z']}:\overline{P}_{[z']}|\tilde{E})=0\ ,
  \ee
which implies the above mentioned impossibility. $\Box$


\subsection{Correlations without secrecy}

In this section, we will characterize those $P_\Omega$ that can be
established by LOPC. This is the content of the following theorem.

\bigskip

{\bf Theorem 6:} {\em A probability distribution $P_\Omega$ can be
generated by LOPC if, and only if, for all bipartite splittings
$\part$,
  \be
  \Omega_{[s]} \geq \Omz
  \label{form}
  \ee
holds.}

\bigskip

{\em Proof:} Let us start by the {\em only if} part. In the proof
of Lemma 3 we have seen that whenever $\Omega_{[s]} \geq \Omz$,
the intrinsic information for the corresponding bipartite
splitting is zero, $I(\P_{[s]} :\overline{\P}_{[s]} \downarrow
E)=0$. It has been proven in \cite{RW} that $I_\downarrow=0$ if,
and only if, $I_{\mbox{\scriptsize form}}=0$. This result and
Lemma 2 imply that \w{form} is a necessary condition for
$P_\Omega$ being generated by LOPC.

For the {\em if} part of the proof, we proceed as follows. First,
we introduce a probability distribution $P_\Omega'$ and prove it
cannot be less secret than $P_\Omega$. This is done by showing
that $P_\Omega'$ can be obtained from $P_\Omega$ by degradating
Eve's information; namely, there exists a map for Eve's random
variable $E\rightarrow \tilde{E}$ such that $P_\Omega\rightarrow
P_\Omega'$. Next, we give an explicit LOPC protocol producing the
probability distribution $P_\Omega'$ without any additional
resource. Thus, $I_{\mbox{\scriptsize form}}$ is zero for
$P_\Omega'$. Then, it follows from the definition of information
of formation that $P_\Omega$ has also $I_{\mbox{\scriptsize
form}}=0$.

With each $P_\Omega$ such that \w{form} holds for all $[s]$, we
associate the following distribution $P'_\Omega$
\begin{center}
\begin{tabular}{|cc|c|c|} \hline
  $A_1\ldots A_{n-1}$ &$A_n$ &$\tilde E$ &$P'_\Omega$ \\
  \hline\hline
  $[0]$ &0 &x &$\Omega_{[0]}$ \\
  $[\bar{0}]$ &1 &x &$\Omega_{[0]}$ \\ \hline
  $[1]$ &0 &x &$\Omega_{[0]}$ \\
  $[\bar{1}]$ &1 &x &$\Omega_{[0]}$ \\
  $[1]$ &0 &$[1]\,0$ &$\Omega_{[1]}-\Omz$ \\
  $[\bar{1}]$ &1 &$[\bar{1}]\,1$ &$\Omega_{[1]}-\Omz$ \\
  &&&\\
  \vdots &\vdots &\vdots &\vdots \\
  &&&\\
  $[s]$ &0 &x &$\Omz$ \\
  $[\bar{s}]$ &1 &x &$\Omz$ \\
  $[s]$ &0 &$[s]\,0$ &$\Om-\Omz$ \\
  $[\bar{s}]$ &1 &$[\bar{s}]\,1$ &$\Om-\Omz$ \\
  &&&\\
  \vdots &\vdots &\vdots &\vdots \\
  &&&\\
  $[2^{n-1}-1]$ &0 &x &$\Omz$ \\
  $[ \overline{2^{n-1}-1} ]$ &1 &x &$\Omz$ \\
  $[2^{n-1}-1]$ &0 &$[2^{n-1}-1]\,0$ &$\Omega_{[2^{n-1}-1]}-\Omz$ \\
  $[ \overline{2^{n-1}-1} ]$ &1 &$[\overline{2^{n-1}-1}]\,1$ &$\Omega_{[2^{n-1}-1]}-\Omz$ \\ \hline
\end{tabular}
\vspace{10pt}
\end{center}
Actually, $P'_\Omega$ can be obtained from $P_\Omega$ after the
following stochastic map on Eve's random variable,
$E\rightarrow\tilde{E}$. If $E$ is equal to ``$[0]0$ or $[\bar
0]1$" we assign $\tilde{E}:=$``x". If $E$ is equal to $[s]\, 0$
($[\bar{s}]\, 1$), we assign $\tilde{E}:=$ ``x" with probability
$\Omz/\Om$, and, with probability $1-\Omz/\Om$ we assign
$\tilde{E}:=E$. Now, we prove that all probability distributions
$P'_\Omega$ of that kind can be created with the following LOPC
protocol. With probability $2^n\, \Omz$ party $\A_1$ broadcasts
the public message ``x" . After receiving ``x", each party
$\A_1\ldots\A_n$ locally generates a random bit $A_1\ldots A_n$.
With probability $\Omega_{[s]}-\Omz$, party $\A_1$ broadcasts the
public message $[s]0$ ($[\bar{s}]1$). After receiving this message
each party outputs its corresponding bit from the sequence $[s]0$
($[\bar{s}]1$). This implies $I_{\mbox{\scriptsize form}}=0$ for
$P'_\Omega$, and the same result applies to $P_\Omega$. $\Box$

\subsection{Construction of the examples}
\label{exconstr}

In this section we explicitly construct the examples that have
been introduced at the beginning of the paper. This is done by
repeatedly using Theorem 5.

{\bf Example 1.} Let us design a probability distribution of $n$
honest parties which is distillable if, and only if, more than $m$
parties cooperate in the protocol. In other words, if $n-m$
parties (or more) do not cooperate, distillation is impossible.
Consider the situation where there are $m$ cooperating parties.
For each bipartition of them, $\P_{[z']}$, there are $2^{n-m}$
different ways of distributing the $n-m$ non-cooperating parties
between the two groups. That is, there are $2^{n-m}$ different
bipartitions $\part$ associated with $\P_{[z']}$. If we set
$\Om=\Omz/2^{n-m}$ for all $[s]\neq [0]$, equation \w{non-coop}
will be satisfied if, and only if, the sum $\sum_{\forall [s]\sim
[z']} \Omega_{[s]}$ has less than $2^{n-m}$ terms, and this only
happens if the number of cooperating parties is larger than $m$.
Notice that $\Omz$ is fixed by the normalization condition
\w{norm}. Because all $\Om$ with $[s]\neq [0]$ have the same
value, distillation is possible even when the cooperating parties
are all separated.

{\bf Example 2.} Let us construct an $n$-party distribution, which
is distillable if, and only if, the cooperating parties join in
groups of at least $k$ people, independently of how many parties
do not cooperate. We denote by $W_{[s]}$ the number of ones that
the bit string $[s]$ has. We impose $\Om=0$ for all strings with
$k \leq W_{[s]} \leq n-k$, and $\Om=\Omz$ for the rest. As before,
$\Omz$ is fixed by the normalization condition \w{norm}. It is
easy to see that, if there is a group of less than $k$ parties,
the bipartition having these $k$ parties in one side and the rest
in the other side, satisfies $\Omega_{[s]}=\Omz$, and this
prevents condition \w{non-coop} from being satisfied. When all
cooperating groups contain at least $k$ people, all bipartitions
that do not split any of the groups satisfy $\Omega_{[s]}=0$, in
this case, condition \w{non-coop} holds independently of how many
parties do not cooperate.

{\bf Example 3.} This $n$-party distribution is distillable if,
and only if,, more than $m$ parties participate in the protocol,
AND, they join in groups of at least $k$ people. This is achieved
with the following assignments. We set $\Om=\Omz/2^{n-m}$ if
$k\leq W_{[s]} \leq n-k$, and $\Om=\Omz$ otherwise. As in the
example 2, if there is one group of less than $k$ cooperating
parties condition \w{non-coop} does not hold. Reasoning in the
same fashion as in example 1, if there are $m$ or less cooperating
parties distillation is impossible.

{\bf Example 4.} This $n$-party distribution is distillable if,
and only if,, there are more than $m$ cooperating parties, OR,
they joint in groups of at least $k$ people, or both. We set
$\Om=0$ if  $k\leq W_{[s]} \leq n-k$, and $\Om=\Omz/2^{n-m}$ for
the rest.

{\bf Example 5.} This $n$-party distribution is distillable if,
and only if, parties $A_i A_j$ cooperate and remain always
together. We suppose without loss of generality that $i,j\neq n$.
If the $i\th$ and $j\th$ bits of the string $[s]$ have the same
value we set $\Om=0$, and $\Om=\Omz$ otherwise. It is clear that
this fulfils our demand. A variation of this example is when
distillation is possible if, and only if, parties $A_i A_j$ remain
separated. The construction of this case is closely analogous to
the previous one. Now, we set $\Om=0$ if $[s]$ has different
values for the $i\th$ and $j\th$ bits, and $\Om=\Omz$ otherwise.

\section{Bound information}
\label{binf}

Bound information represents the cryptographic analog of bound
entanglement, an intriguing feature of some quantum states found
by the Horodeckis in 1998 \cite{Horo}. In the bipartite case,
bound information can easily be defined using the previously
introduced quantities \cite{GW}: a probability distribution
$P(A,B,E)$ contains bound information when the following two
conditions hold: \bea
  S&=&0 \nonumber\\
  I_{\mbox{\scriptsize form}}&>&0\ .
\eea Therefore, $P(A,B,E)$ has bound information when (i) no
secret-key bits can be extracted from it by LOPC, but (ii) its
formation by LOPC is impossible. In other words, the non-zero
secrecy content of the probability is bound because secret
correlations are necessary for its preparation but cannot be
distilled into a pure form. There exist several results supporting
the existence of this analog of bound entanglement in the
bipartite case: in Refs. \cite{GRW,GW} several probability
distributions were constructed for which one can prove that
$I_\downarrow$ is strictly positive but none of the known
secret-key distillation protocols allow to extract secret bits.
Moreover, it was shown in \cite{RW} that there exist probability
distribution where $S<I_{\mbox{\scriptsize form}}$. This already
proves the irreversibility, in terms of secret bits, in the
processes of formation and key distillation for some probability
distributions. Actually, the authors of \cite{RW} constructed a
family of probability distributions where $I_{\mbox{\scriptsize
form}}>1/2$ while $S$ can be arbitrarily small. Unfortunately, no
example of $P(A,B,E)$ such that $0=S<I_{\mbox{\scriptsize form}}$
is known until now.

Bound information was initially defined in the case of two honest
parties. However, its generalization to the multipartite scenario
is again straightforward: a probability distribution $P(A_1,\dots
,A_n,E)$ has bound information when (i) its formation by LOPC is
impossible and (ii) no secret-key bits can be extracted between
any pair of parties by LOPC. In what follows, we use the
techniques described above in order to show the existence of
multipartite bound information. We will do that for the case of
three honest parties. Moreover, we will see that similarly to what
happens in the quantum case, bound information can be activated:
the combination of different probability distributions with bound
information may give a distillable probability distribution.

\subsection{Proof of the existence of bound information}
\label{binfpr}

In this section we prove the existence of bound information in the
tripartite scenario. In order to do that we give a probability
distribution $P(A_1,A_2,A_3,E)$ and show that its formation by
LOPC is impossible but no secret-key bits can be extracted from it
by the honest parties using LOPC. Although these results already
appear in \cite{us}, here we review them using the formalism
described in the previous section.

Using the introduced notation, an example of tripartite
probability distribution having bound information reads as
follows:
\begin{center}
\begin{tabular}{|cc|c|c|} \hline
  $A_1A_2$ &$A_3$ &$E$ &$P_1$ \\
  \hline\hline
  $[0]$ &0 &$[0]0$ or $[\bar{0}]1$ &$1/6$ \\
  $[\bar{0}]$ &1 &$[0]0$ or $[\bar{0}]1$ &$1/6$ \\ \hline
  $[1]$ &0 &$[1]0$ &$1/6$ \\
  $[\bar 1]$ &1 &$[\bar 1]1$ &$1/6$ \\
  $[2]$ &0 &$[2]0$ &$0$ \\
  $[\bar 2]$ &1 &$[\bar 2]1$ &$0$ \\
  $[3]$ &0 &$[3]0$ &$1/6$ \\
  $[\bar 3]$ &1 &$[\bar 3]1$ &$1/6$ \\ \hline
\end{tabular}
\vspace{10pt}
\end{center}
Note that the role played by $A_2$ and $A_3$ in $P_1$ is the same,
up to a relabelling of Eve's variables.

Using Lemmas 1 and 3, it is relatively simple to see that no pair
of parties can distill secret bits from this probability
distribution. Consider, for instance, the partition
$A_2-(A_1A_3)$, corresponding to $[s]=[1]=01$. Because of Lemma 3,
no distillation is possible since $\Omega_{[1]}=\Omega_{[0]}=1/6$.
Then, Lemma 1 implies that $A_2$ can distill secret bits neither
with $A_1$ nor with $A_3$. Because of the symmetry of the
distribution, the same result holds for $(A_1A_2)-A_3$. Therefore
no pair of parties can distill a secret key. Finally, consider the
third partition $A_1-(A_2A_3)$. In this case, where $[s]=[2]=10$,
we have $\Omega_{[2]}=0<\Omega_{[0]}=1/6$. That is, the
probability distribution corresponding to this partition is
distillable, which means that it could not have been created by
LOPC. Using Lemma 2, this implies that the initial probability
distribution $P_1$ cannot be created by LOPC either. This proves
that the non-zero secrecy content of $P_1$ is bound, i.e. it
constitutes an example of bound information.

The proof presented here is almost the same as in \cite{us},
having been adapted to the notation introduced above. Actually,
the non-distillability of $P_1$ could alternatively have been
proven using Lemma 4. Note also that many of the probability
distributions given above, such as Example 2, already constituted
examples of bound information.

\subsection{Bound information can be activated}

The activation of bound entanglement is perhaps one of the most
surprising results found in entanglement theory \cite{DC,SST}.
Bound entanglement is said to be activated whenever one can
distill pure-state entanglement from the combination of several
bound entangled states. Remarkably, in some cases this activation
can be achieved by mixing different bound entangled states
\cite{DC2}! As it will be shown shortly, a similar feature is
observed for classical probability distributions.

Consider the situation where three honest parties and an
eavesdropper have access to correlated random variables described
by $P_1$. In addition, they also have access to other random
variables described by $P_2$ and $P_3$, where these two
probability distributions correspond to cyclic permutations,
$A_1\rightarrow A_2\rightarrow A_3\rightarrow A_1$, of $P_1$. Of
course, $P_1$, $P_2$ and $P_3$ have bound information. Now, the
three honest parties forget what the actual distribution is.
Alternatively, one can think that a source is sending to the
parties random variables correlated through $P_1$, $P_2$ and $P_3$
with equal probability, and the information about the prepared
probability distribution is only accessible to Eve. The resulting
distribution, $P_{\mbox{\scriptsize res}}$, can be described as

\begin{center}
\begin{tabular}{|cc|c|c|} \hline
  $A_1A_2$ &$A_3$ &$E$ &$P_{\mbox{\scriptsize res}}$ \\
  \hline\hline
  $[0]$ &0 &$[0]0$ or $[\bar{0}]1$ &$1/6$ \\
  $[\bar{0}]$ &1 &$[0]0$ or $[\bar{0}]1$ &$1/6$ \\ \hline
  $[1]$ &0 &$[1]0$ &$1/9$ \\
  $[\bar 1]$ &1 &$[\bar 1]1$ &$1/9$ \\
  $[2]$ &0 &$[2]0$ &$1/9$ \\
  $[\bar 2]$ &1 &$[\bar 2]1$ &$1/9$ \\
  $[3]$ &0 &$[3]0$ &$1/9$ \\
  $[\bar 3]$ &1 &$[\bar 3]1$ &$1/9$ \\ \hline
\end{tabular}
\vspace{10pt}
\end{center}

It is now straightforward to see that this probability
distribution is distillable, even in the fully multipartite
scenario where the three parties remain separated. This follows
from Lemma 4, since for all the partitions one has $1/9<1/6$.
Therefore, the combination of non-distillable probability
distributions produces a distillable distribution.

\section{Conclusions}

In this work, we have presented a family of probability
distributions in the multipartite scenario of $n$ honest parties
and an eavesdropper. Using this family, we were able to construct
different examples of probability distribution with a huge variety
of secrecy properties. This rich variety of examples shows how
intricate the structure of multipartite secret correlations is.
Moreover, the introduced techniques allowed us to prove the
existence and activation of bound information, namely
non-distillable secret correlations.

We would like to mention here some analogies between our results
and the problem of entanglement manipulations in Quantum
Information Theory (see Refs. \cite{GW,CP}). The intuition for the
construction of the previous family of probability distributions
came from the quantum states discussed in Refs. \cite{DC,DC2}.
Indeed, these distributions represent the cryptographic classical
analog of these states. Moreover, the existence of bound
information, that was our initial motivation for this study, was
conjectured in 2000 \cite{GW} as a classical counterpart of bound
entanglement. In this sense, all these results constitute one of
the first examples where well-established ideas in Quantum
Information Theory have successfully been translated to the
classical side. Up to now, the flow of results has mainly been in
the opposite direction.

Unfortunately, the existence of bipartite bound information, that
is, probability distributions with non-distillable secret
correlations, remains as an open question. Indeed, the existence
of multipartite bound information exploited the possibility of
considering splittings of the parties into different groups,
something impossible in the bipartite scenario. In this sense, it
is an interesting issue to study how those quantum concepts that
allowed to prove the existence of bound entanglement for quantum
states, such as partial transposition \cite{Peres}, can be adapted
to the key-agreement scenario.

\section*{Appendix: Repeated code protocol}

In this appendix, the repeated code protocol is described.
Consider $m$ separated parties $\A_1,\ldots,\A_m$ willing to
generate an $m$-partite secret key. Each of these parties, say
$\A_i$, has access to many realizations of its corresponding
random variable $A_i$. Additionally, there is an eavesdropping
party, Eve, who has access to a random variable $E$ correlated to
$A_i$ through the probability distribution $P\!\left( A_1,\ldots
,A_m,E\right)$. Note that it is assumed that each realization of
$A_1,A_2,\ldots ,A_m,E$, is independent of the other. Moreover,
this probability distribution is known by all the parties.

The first part of this key distillation protocol is implemented by
the following three steps:
\begin{enumerate}
\item Each party takes $N$ realizations of her own random
variable:
  \bea
  &A_1^{N}& =\left( A_1^{(1)},A_1^{(2)},\ldots ,A_1^{(N)}
  \right) \nonumber\\
  &\vdots& \nonumber\\
  &A_m^{N}& =\left( A_m^{(1)},A_m^{(2)},\ldots ,A_m^{(N)}
  \right) \nonumber
  \eea

\item One of the honest parties ---say $\A_1$--- generates locally
a random bit $k_1$, computes the $N$ numbers $X_r:=(k_1+A_1^{(r)}
\bmod 2)$ for $r=1,\ldots ,N$, and broadcasts through the public
channel the $N$-bit string:
  \be
  \left( X_1, X_2, \ldots ,X_N \right)\ .
  \label{string}
  \ee

\item All the remaining parties ---in this case $\A_2,\ldots
,\A_n$--- perform the following operation. Party $A_i$ adds
bitwise the broadcasted string \w{string} to his symbols
$(A_i^{(1)},A_i^{(2)},\ldots ,A_i^{(N)})$. If he obtains the same
result for all of them, that is $(X_r+A_i^{(r)} \bmod 2)=k_i$ for
$r=1,\ldots ,N$, he accepts $k_i$ and communicates the acceptance
to the other parties. If not, all parties reject the $N$
realizations of $P_{\Omega}$.
\end{enumerate}

The final step to attain a secret key uses as input many
realizations of $(k_1,\ldots ,k_m)$. It consists of the one-way
distillation protocol given by Csisz\'ar and K\"orner in
\cite{CK}. The fact that this protocol is designed for two parties
is not a problem. In our case, one of the honest parties, say
$\A_1$, broadcasts all public messages to the rest, who perform
error correction and privacy amplification to their data.

Let us analyze under which conditions a probability distribution
belonging to the family $P_\Omega$ can be distilled into a secret
key
using this protocol. 
In the usual notation, a probability distribution $P_\Omega$ reads
\begin{center}
\begin{tabular}{|cc|c|c|} \hline
  $A_1\ldots ,A_{m-1}$ &$A_m$ &$E$ &$P_\Omega$ \\
  \hline\hline
  $[0]$ &0 &$[0]0$ or $[\bar{0}]1$ &$\Omega_{[0]}$ \\
  $[\bar{0}]$ &1 &$[0]0$ or $[\bar{0}]1$ &$\Omega_{[0]}$ \\ \hline
  &&&\\
  \vdots &\vdots &\vdots &\vdots \\
  &&&\\
  $[s]$ &0 &$[s]\,0$ &$\Om$ \\
  $[\bar{s}]$ &1 &$[\bar{s}]\,1$ &$\Om$ \\
  &&&\\
  \vdots &\vdots &\vdots &\vdots \\
  &&&\\\hline
\end{tabular}
\vspace{10pt}
\end{center}
%
Notice that when party $\A_i$, with $i\geq 2$, accepts $k_i$ in
step 3, the variables $A_i^{(1)},\ldots ,A_i^{(N)}$ are all equal,
or, all different to $A_1^{(1)},\ldots ,A_1^{(N)}$.
%
%
%
This is equivalent to saying that, the $N$ realizations of
$P_\Omega$ used in this first part of the protocol, have to be all
in the same pair of events, characterized by a given $s_0$, that
is, $A_1^{(r)},\ldots,A_m^{(r)}=[s_0]0$ or
$A_1^{(r)},\ldots,A_m^{(r)}=[\bar s_0]1$ for $r=1,\ldots,N$. This
happens with probability
  \be
  p\,(s_0)=\frac{\Omega_{[s_0]}^N}
  {\sum_{[s]}\Omega_{[s]}^N}\ .
  \label{case1}
  \ee
Notice that in the case $[s_0]=[0]$, the bits $k_1,\ldots,k_m$ are
all equal, and Eve has no knowledge about them. If $\Omz >
\Omega_{[s]}$ for all $[s]\neq [0]$, the probability $p(0)$ tends
to one when making $N$ large. Thus, choosing a large enough $N$,
the honest parties can obtain a probability distribution that can
be distilled to secret key with non-zero rate by means of the
one-way reconciliation techniques of \cite{CK}.


{\bf Result:} {\em An $m$-partite distribution of the family
$\P_\Omega$ can be distilled if $\Omega_{[0]}>\Omega_{[s]}$ for
all $[s]\neq[0]$.}

\section*{Acknowledgment}
The authors would like to thank Ignacio Cirac, Nicolas Gisin, Nick
Jones, Renato Renner and Stefan Wolf for discussion. This work has
been supported by the the U.K. Engineering and Physical Sciences
Research Council (IRC QIP), the Spanish Ministerio de Ciencia y
Tecnolog\'\i a, under the ``Ram\'on y Cajal" grant, and the
Generalitat de Catalunya.


%

\end{document}